\documentclass[aps,prc,twocolumn,superscriptaddress,preprintnumbers,amsmath,amssymb,floatfix,showkeys,nofootinbib]{revtex4}

\usepackage{graphicx}
\usepackage{currfile}
\usepackage{epsfig}
\usepackage{dcolumn}
\usepackage{bm}
\usepackage{longtable}
\usepackage{color}
\usepackage{float}
\usepackage{comment}
\usepackage{amsmath}
\usepackage{amsfonts}
\usepackage{xspace}
\usepackage{multirow}
\usepackage{nicefrac}
\usepackage{xspace}
\usepackage[normalem]{ulem}
\usepackage{siunitx}

\usepackage{array}
\newcolumntype{L}[1]{>{\raggedright\let\newline\\\arraybackslash\hspace{0pt}}m{#1}}
\newcolumntype{C}[1]{>{\centering\let\newline\\\arraybackslash\hspace{0pt}}m{#1}}
\newcolumntype{R}[1]{>{\raggedleft\let\newline\\\arraybackslash\hspace{0pt}}m{#1}}

\makeatletter
\g@addto@macro\bfseries{\boldmath}
\makeatother

\usepackage{tikz}
\def\checkmark{\tikz\fill[scale=0.4](0,.35) -- (.25,0) -- (1,.7) -- (.25,.15) -- cycle;} 

\begin{document}

\newcommand{\ChiEFT}{\ensuremath{\chi}EFT\xspace}
\newcommand{\PiM}{$\pi^-$\xspace}
\newcommand{\PiP}{$\pi^+$\xspace}
\newcommand{\be}{\begin{equation}}
\newcommand{\ee}{\end{equation}}
\newcommand{\geant} {{{G}\texttt{\scriptsize{EANT}}4}}

\preprint{MAX-lab $\gamma d \rightarrow \pi^-pp$ Summary article for PRC}

\title{Near-threshold $\pi^-$ photoproduction on the deuteron}


%
%

\author{\mbox{B.~Strandberg}}
\affiliation{School of Physics and Astronomy, University of Glasgow, Glasgow G12 8QQ, Scotland UK}

\author{\mbox{K.~G.~Fissum}}
  \altaffiliation{Corresponding author; \texttt{kevin.fissum@nuclear.lu.se}}
\affiliation{Department of Physics, Lund University, SE-221 00 Lund, Sweden}

\author{\mbox{J.~R.~M.~Annand}}
\affiliation{School of Physics and Astronomy, University of Glasgow, Glasgow G12 8QQ, Scotland UK}

\author{\mbox{W.~J.~Briscoe}}
\affiliation{Institute for Nuclear Studies, Department of Physics, The George Washington University, Washington DC 20052, USA}

\author{\mbox{J.~Brudvik}}
\affiliation{MAX IV Laboratory, Lund University, SE-221 00 Lund, Sweden}

\author{\mbox{F.~Cividini}}
\affiliation{Institut f{\"ur} Kernphysik, Johannes Gutenberg-Universit{\"at} Mainz, D-55099 Mainz, Germany}

\author{\mbox{L.~Clark}}
\affiliation{School of Physics and Astronomy, University of Glasgow, Glasgow G12 8QQ, Scotland UK}

\author{\mbox{E.~J.~Downie}}
\affiliation{Institute for Nuclear Studies, Department of Physics, The George Washington University, Washington DC 20052, USA}

\author{\mbox{K.~England}}
\affiliation{College of Engineering, University of Massachusetts Dartmouth, North Dartmouth MA 02747, USA}

\author{\mbox{G.~Feldman}}
\affiliation{Institute for Nuclear Studies, Department of Physics, The George Washington University, Washington DC 20052, USA}

\author{\mbox{D.~I.~Glazier}}
\affiliation{School of Physics and Astronomy, University of Glasgow, Glasgow G12 8QQ, Scotland UK}

\author{\mbox{K.~Hamilton}}
\affiliation{School of Physics and Astronomy, University of Glasgow, Glasgow G12 8QQ, Scotland UK}

\author{\mbox{K.~Hansen}}
\affiliation{MAX IV Laboratory, Lund University, SE-221 00 Lund, Sweden}

\author{\mbox{L.~Isaksson}}
\affiliation{MAX IV Laboratory, Lund University, SE-221 00 Lund, Sweden}

\author{\mbox{R.~Al~Jebali}}
\affiliation{School of Physics and Astronomy, University of Glasgow, Glasgow G12 8QQ, Scotland UK}

\author{\mbox{M.~A.~Kovash}}
\affiliation{Department of Physics and Astronomy, University of Kentucky, Lexington, KY 40506, USA}

\author{\mbox{A. E. Kudryavtsev}}
\affiliation{Institute for Nuclear Studies, Department of Physics, The George Washington University, Washington DC 20052, USA}
\affiliation{National Research Centre ``Kurchatov Institute", Institute for Theoretical and Experimental Physics (ITEP), Moscow 117218, Russia}

\author{\mbox{V.~Lensky}}
\affiliation{National Research Centre ``Kurchatov Institute", Institute for Theoretical and Experimental Physics (ITEP), Moscow 117218, Russia}
\affiliation{Institut f\"ur Kernphysik \& Cluster of Excellence PRISMA, Johannes Gutenberg Universitat, Mainz D-55099, Germany}

\author{\mbox{S.~Lipschutz}}
\affiliation{Institute for Nuclear Studies, Department of Physics, The George Washington University, Washington DC 20052, USA}
\altaffiliation{Present address: National Superconducting Cyclotron Laboratory, Michigan State University, Michigan 48824, USA}

\author{\mbox{M.~Lundin}}
\affiliation{MAX IV Laboratory, Lund University, SE-221 00 Lund, Sweden}

\author{\mbox{M.~Meshkian}}
\affiliation{Department of Physics, Lund University, SE-221 00 Lund, Sweden}

\author{\mbox{D.~G.~Middleton}}
\affiliation{Institut f{\"ur} Kernphysik, Johannes Gutenberg-Universit{\"at} Mainz, D-55099 Mainz, Germany}
\affiliation{Mount Allison University, Sackville, New Brunswick E4L 1E6, Canada.}

\author{\mbox{L.~S.~Myers}}
\affiliation{Bluffton University, 1 University Drive, Bluffton, OH 45817, USA}

\author{\mbox{D.~O'Donnell}}
\affiliation{School of Physics and Astronomy, University of Glasgow, Glasgow G12 8QQ, Scotland UK}

\author{\mbox{G.~V.~O'Rielly}}
\affiliation{Department of Physics, University of Massachusetts Dartmouth, Dartmouth MA 02747, USA}

\author{\mbox{B.~Oussena}}
\affiliation{Institute for Nuclear Studies, Department of Physics, The George Washington University, Washington DC 20052, USA}

\author{\mbox{M.~F.~Preston}}
\affiliation{Department of Physics, Lund University, SE-221 00 Lund, Sweden}

\author{\mbox{B.~Schr\"oder}}
\affiliation{Department of Physics, Lund University, SE-221 00 Lund, Sweden}
\affiliation{MAX IV Laboratory, Lund University, SE-221 00 Lund, Sweden}

\author{\mbox{B.~Seitz}}
\affiliation{School of Physics and Astronomy, University of Glasgow, Glasgow G12 8QQ, Scotland UK}

\author{\mbox{I.~I.~Strakovsky}}
\affiliation{Institute for Nuclear Studies, Department of Physics, The George Washington University, Washington DC 20052, USA}

\author{\mbox{M.~Taragin}}
\affiliation{Institute for Nuclear Studies, Department of Physics, The George Washington University, Washington DC 20052, USA}

\author{\mbox{V.~E.~Tarasov}}
\affiliation{National Research Centre ``Kurchatov Institute", Institute for Theoretical and Experimental Physics (ITEP), Moscow 117218, Russia}

\collaboration{The PIONS@MAX-lab Collaboration}
\noaffiliation

\date{\today}


\begin{abstract}

  The first experimental investigation of the near-threshold cross section for incoherent \PiM photoproduction on the deuteron $\gamma  d \rightarrow \pi^- pp$ is presented. The experimental technique involved detection of the $\sim$131~MeV gamma ray resulting from the radiative capture of photoproduced \PiM in the target. The total cross section has been measured using an unpolarized tagged-photon beam, a liquid-deuterium target, and three very large NaI(Tl) spectrometers. The data are compared to theoretical models that give insight into the elementary reaction $\gamma n \rightarrow \pi^- p$ and pion-nucleon and nucleon-nucleon final-state interactions.
  

\end{abstract}

\maketitle

\section{Introduction}
Incoherent pion photoproduction on the deuteron $\gamma  d \rightarrow \pi\!N\!N$ provides information on the elementary reaction on the nucleon $\gamma N\rightarrow \pi N$ and on pion-nucleon ($\pi\!N$) and nucleon-nucleon ($N\!N$) final-state interactions (FSI). The near-threshold cross section for the elementary reaction is sensitive to the $E_{0+}$ amplitude, which has a long history of theoretical studies closely related to measurements of near-threshold pion photoproduction~\cite{Drechsel:1992pn}. Partial-wave analysis (PWA)~\cite{PhysRevC.42.1853} of experimental data sets may be used to obtain values for this and other photoproduction amplitudes. These are vital inputs to low-energy descriptions of hadron physics based on dispersion relations~\cite{Drechsel2007} or chiral perturbation theory ($\chi$PT)~\cite{PhysRevC.88.055207}. The latter, which is also used for comparison with experimental data in this article, is an effective field theory of Quantum Chromodynamics (QCD), where hadrons, instead of quarks and gluons, act as relevant degrees of freedom. $\chi$PT emerges from the QCD Lagrangian in the chiral limit of vanishing up and down quark masses $(m_u,\:m_d \rightarrow 0)$ and thus offers a way to investigate the fundamental symmetries and interactions of the strong force in an energy regime where QCD is non-perturbative.

Tagged-photon beams combined with improved detector technology have substantially increased the size of the global pion-production data set over the last decades. However, most measurements have focused on the $\pi^0$ channel~\cite{PhysRevC.53.R1052, PhysRevC.55.2016, PhysRevC.58.2574, PhysRevLett.111.062004}, as the elementary amplitude for $\pi^0$ production vanishes in the chiral limit. Thus the $\pi^0$ data allow for direct probing of chiral symmetry breaking phenomena. The most recent of these experiments~\cite{PhysRevLett.111.062004} provided high-precision differential cross section and beam asymmetry data that have enabled stringent testing of $\chi$PT. Threshold measurements of charged pion photoproduction are scarce in comparison. While the threshold cross section for $\pi^+$ photoproduction was established in Ref.~\cite{Booth_1979}, none of the $E_\gamma < 200$~MeV \PiM measurements~\cite{Rossi1973, SALOMON1984493, Wang1992, Liu1994} have probed the near-threshold region, with the lowest-energy data point at ${\sim}158$~MeV~\cite{Liu1994}, more than 10~MeV above threshold. This article reports the pioneering measurement of the total cross section for \PiM photoproduction on the deuteron in the energy range 147~--~160~MeV. The well-understood radiative capture (RC) reaction on the deuteron $\pi^- d \rightarrow \gamma nn$~\cite{gabioud1979} with an end-point photon-energy of 131.4~MeV is exploited in a novel way for the yield determination of the photoproduced \PiM.

\section{Experimental setup}
The experiment was performed at the Tagged-Photon Facility~\cite{adler2012} of the MAX IV Laboratory~\cite{eriksson2014} in Sweden. A tagged-photon beam with energies from ${140-160}$~MeV, created via the bremsstrahlung-tagging technique~\cite{adler1990,adler1997}, was incident on a thin cylindrical Kapton vessel that contained liquid deuterium (LD$_2$) with density ${\rho_D = (0.163 \pm 0.001)}$~g/$\rm cm^3$. The Kapton vessel was a cylinder of 170~mm length and 68~mm diameter, aligned along the axis of the photon beam. The vessel walls were 120~$\mu$m thick. The tagged-photon energies $E_\gamma$ were determined by momentum analysis of the post-bremsstrahlung electrons using a dipole magnet together with a 64-channel focal-plane (FP) hodoscope~\cite{vogt1993}. The tagged-photon energy resolution was ${\pm}0.3$~MeV. Electron arrival times at the hodoscope were digitized with multi-hit time-to-digital converters (TDCs). The post-bremsstrahlung electron counting rate (typically $0.1-1$~MHz per FP channel), necessary for the photon-flux determination, was measured by scalers, normalized to the counting time. Tagging efficiency, the fraction of bremsstrahlung photons which passed through the photon-beam collimation system en route to the target, was measured daily. The mean tagging efficiency was $({\sim}23\pm2^{\rm sys.})$\%.

\begin{figure}[t]
    \centering
    \includegraphics[scale=0.58]{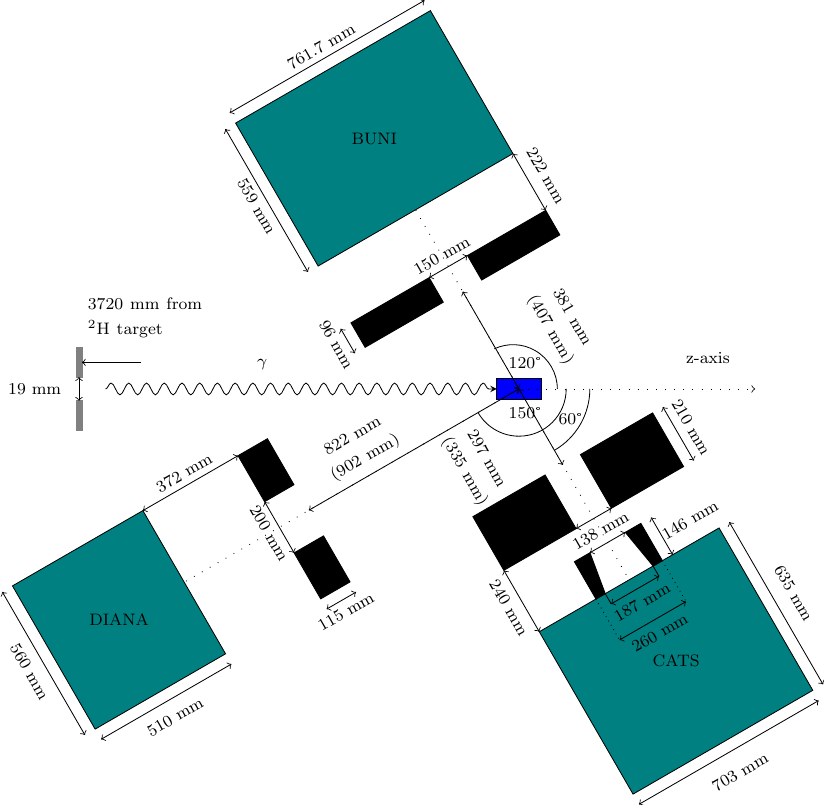}
    \caption{(Color online) A schematic plan view of the experimental setup. Gray - beam collimator; wavy line - photon beam; blue box - deuterium target; black regions - detector front and inner collimators; green boxes - scintillators (NaI(Tl) and plastics). The distances between the target and the detectors were different in 2011 and 2015. The 2011 values are indicated in brackets.}
    \label{fig:floorplan}
\end{figure}

Three large NaI(Tl) spectrometers, named Boston University Sodium Iodide (BUNI)~\cite{miller1988}, Compton and Two Photon Spectrometer (CATS)~\cite{Hunger1997385} and Detector Of Iodine And Sodium (DIANA)~\cite{myers2010}, were placed at laboratory angles $\theta = 60^\circ,\:120^\circ$ and $150^\circ$ to detect RC photons originating from the LD$_2$ target. The positioning of the detectors relative to the beam and the target is depicted in Fig.~\ref{fig:floorplan}. Each spectrometer consisted of a cylindrical core crystal surrounded by an annulus of optically isolated crystal segments. The segments were in turn surrounded by plastic scintillators. Scintillation light was read out by photomultiplier tubes (PMTs) attached to the rear faces of the scintillators. Analog signals from the PMTs were recorded by charge-integrating analog-to-digital converters (ADCs).

Data were recorded on an event-by-event basis. The data-acquisition and data-analysis software were based on ROOT~\cite{Brun199781} and RooFit~\cite{Verkerke:2003ir} frameworks. The data acquisition was triggered by an energy deposition greater than ${\sim}50$~MeV in any NaI(Tl), which initiated the readout of the ADCs and started the TDCs. The TDC stop signals came from the post-bremsstrahlung electrons striking the FP channels. The ADC information was used to reconstruct detected photon energies, whereas the FP TDC information established the coincidence between the post-bremsstrahlung recoil electrons and the particles detected with the spectrometers. The data were collected over three run periods in 2011 and 2015.

\section{Analysis}

\subsection*{Calibration}

\begin{figure}[t]
    \centering
    \includegraphics[scale=1.2]{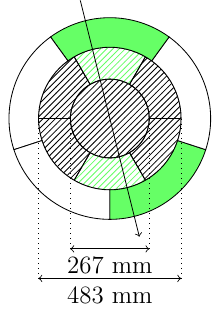}
    \caption{(Color online) Cross-sectional view of the CATS detector. Cosmic-ray muon events (downgoing arrow) that caused a signal in opposing annulus segments (green) were selected for monitoring PMT gain instabilities. NaI(Tl) crystals are striped to distinguish them from plastic scintillators~\cite{PhysRevC.98.012201}.}
    \label{fig:cosmics}
\end{figure}

Each NaI(Tl) detector was calibrated from its in-beam response to a low-intensity tagged-photon beam. Cosmic-ray muons that traversed the detectors during data taking were identified with the annulus scintillators by requiring coincident signals in opposing annular segments. Selection of the cosmic-ray events is illustrated in Fig.~\ref{fig:cosmics}. Shifts in the pulse-height distributions of selected cosmic-ray muon events were used to correct for PMT gain instabilities. After calibration, the NaI(Tl) detectors had a resolution of ${\sim}2$\% (full width at half maximum) for the incident photon energies. The absolute calibration of the tagged-photon energies and the NaI(Tl) detectors was determined with an accuracy of $\pm0.4$~MeV by reconstructing the 131.4~MeV photon-energy end-point from the RC reaction $\pi^- d \rightarrow \gamma nn$~\cite{gabioud1979}.

\subsection*{Signal identification}

\begin{figure}[t]
    \centering
    \includegraphics[scale=0.45]{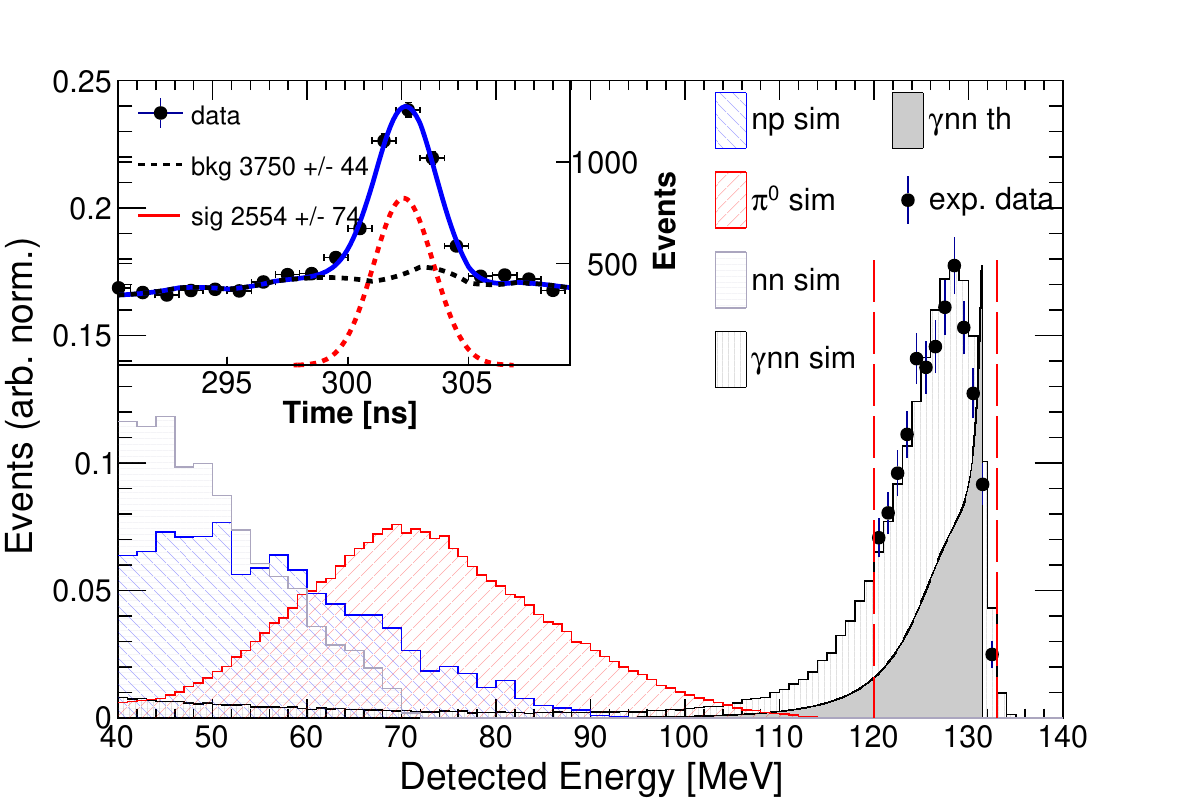}
    \caption{(Color online) Simulated energy spectra of dominant reaction channels alongside the measured energy spectrum and a theoretical energy spectrum of \PiM RC. Inset: A typical fit to a timing-coincidence spectrum for events inside the cut $E_{\rm det} \in [120,\:133]$~MeV for yield determination.}
    \label{fig:espec_plot}
\end{figure}

Photons from the RC reaction were also used to determine the yield of photoproduced \PiM. The near-threshold \PiM had low kinetic energies and most were instantaneously captured inside the LD$_2$ target. The two dominant capture channels are non-radiative capture (NRC) $\pi^- d \rightarrow nn$ (absolute branching ratio ${B\!R_{\rm nrc} = 0.739 \pm 0.010}$) and RC (${B\!R_{\rm rc} = 0.261 \pm 0.004}$)~\cite{Highland1981333}. Using these branching ratios and the energy spectrum of the RC photons~\cite{gabioud1979, PhysRevC.11.90, PhysRevC.16.1976}, the \PiM photoproduction yield was obtained. Figure~\ref{fig:espec_plot} depicts the simulated energy spectra of the dominant background reactions of deuteron photodisintegration (\emph{np~sim}), $\pi^0$ photoproduction (\emph{$\pi^0$~sim}) and \PiM NRC (\emph{nn~sim}), alongside the theoretical RC spectrum (\emph{$\gamma nn$~th})~\cite{PhysRevC.11.90, GibbsGibson}, the simulated RC spectrum (\emph{$\gamma nn$~sim}) and the measured energy spectrum (\emph{exp.~data})\label{text:fig_description}. Simulations were based on \geant~\cite{Geant4}. The photoproduced \PiP did not constitute a significant background, as the muons from the dominant subsequent decay ${\pi^+ \rightarrow \mu^+ \nu_\mu}$ did not deposit more than ${\sim}50$~MeV in any of the NaI(Tl) detectors. Positrons from the decay ${\mu^+\rightarrow e^+ \nu_e \bar{\nu}_\mu}$ were almost always outside the timing coincidence window with respect to the post-bremsstrahlung electron. The simulated RC spectrum was obtained by first matching the Monte Carlo in-beam data to the experimental in-beam data~\cite{PhysRevC.98.012201, PhysRevC.92.025203}. Then, photons with energies sampled from the theoretical RC spectrum and an isotropic angular distribution were generated in the LD$_2$ target into $4\pi$ solid angle. Energy deposited by the photons in the NaI(Tl) detectors was smeared to account for the previously determined resolution effects, which led to the simulated RC spectrum (Fig.~\ref{fig:espec_plot}). The simulation, which is in excellent agreement with the data, indicated that the dominant background reactions could be removed by selecting the detected energy $E_{\rm det} \in [120,\:133]$~MeV. Background from elastic $\gamma d \rightarrow \gamma d$ and inelastic $\gamma d \rightarrow \gamma np$ Compton scattering could not be separated. Contamination from Compton scattering channels was angle- and energy-dependent, but at the present energies, the scattering cross section is only a few percent of the charged-pion photoproduction cross section. The cross-section data from Refs.~\cite{PhysRevC.98.012201,PhysRevLett.88.162301} were extrapolated to produce conservative scattering-contamination estimates. These indicated that the effect on the extracted \PiM cross section was typically $\pm3$\% (maximum of 5.5\% at lowest $E_\gamma$). This effect was accounted for in the systematic uncertainty analysis discussed below.

\subsection*{Yield determination}

The total cross section for \PiM photoproduction on the deuteron was determined according to
\begin{equation}
\sigma = \frac{4\pi Y}{\Omega_{\rm eff} N_\gamma \kappa_{\rm eff} P_{\rm c} B\!R_{\rm rc}}, \label{eq:xsec}
\end{equation}
where $Y$ is the yield of RC photons, $\Omega_{\rm eff}$ is the detector acceptance, $N_\gamma$ is the tagged-photon flux incident on the target, $\kappa_{\rm eff}$ is the effective target thickness, $P_{\rm c}$ is the \PiM capture probability inside the LD$_2$ target and $B\!R_{\rm rc}$ is the branching ratio for RC. The factor $4\pi$ originates from the assumption that RC photons are emitted isotropically. For the yield determination, timing-coincidence spectra with respect to the post-bremsstrahlung recoil electrons were filled for events inside the cut $E_{\rm det} \in [120,\:133]$~MeV. The FP channels were grouped in eight ${\sim}2.5$~MeV wide bins, resulting in eight spectra per detector. The resulting spectra had a coincidence peak superimposed upon events that were in random coincidence. As the dominant background reactions were removed by the cut on $E_{\rm det}$, \PiM capture yields could be determined directly from fits to the coincidence spectra (Fig.~\ref{fig:espec_plot} inset). The signal peak was represented by a Gaussian. The background from random coincidences had a time structure due to a time modulation of the electron-beam intensity related to the pulse-stretching and beam-extraction apparatus~\cite{jebali2013}. The first two FP energy bins were below \PiM/\PiP threshold. Thus, the coincidence spectra for these bins were completely dominated by random coincidences, which allowed estimation of the random-background shape. The background shape visible in the inset of Fig.~\ref{fig:espec_plot} (black line) was obtained from the sub-pion-threshold data and employed in the fit of the super-pion-threshold data. Tools in the RooFit package enable creation of a fit shape from any histogram, which circumvents the difficulty of defining an analytical form for the non-trivial shape of the random background coincidences. The fit was moderately dependent on the width of the fitted window around the coincidence peak, which led to a systematic uncertainty of ${\sim}2$\% (7\% at lowest $E_\gamma$). Systematic uncertainty due to contamination from \PiM produced in the thin-walled Kapton vessel was estimated to be ${\sim}1.5$\% by taking into account the chemical composition of Kapton, the thickness of the endcaps of the vessel and assuming conservatively that the \PiM photoproduction cross section on $\rm {}^{12}C$ and $\rm {}^{16}O$ scales linearly with the number of neutrons per atom.

\subsection*{Detector acceptances}

The detector acceptance $\Omega_{\rm eff}$ was determined from the simulated RC spectrum described previously. The detector acceptance was determined by
\begin{equation}
  \Omega_{\rm eff} = 4\pi N_{E_{\rm det} \in [120,\:133]\:{\rm MeV}}/N_{\rm tot}, \label{eq:acceptance}
\end{equation}
where $N_{\rm tot}$ is the total number of Monte-Carlo photons simulated inside the target, with energies sampled from the theoretical RC spectrum and directions sampled from a phase-space distribution over $4\pi$ solid angle. The numerator is the number of events in a detector within the energy cut $E_{\rm det} \in [120,\:133]$~MeV. The acceptances of the detectors at $60^\circ$, $120^\circ$ and $150^\circ$ were ${\sim}46$~msr, ${\sim}30$~msr and ${\sim}26$~msr, respectively. The dominant systematic uncertainty of 5\% originated from the uncertainty in the theoretical model for RC~\cite{GibbsGibson}. Systematic uncertainty from the positioning accuracy of the detectors and the target was estimated to be ${\sim}3$\% by varying the detector and target positions in the simulation within realistic limits. The $\pm 0.4$~MeV uncertainty in the overall energy calibration of the detectors propagated into the acceptance calculation and was estimated to have an effect of ${\sim}1.5$\% by varying the energy cut by the uncertainty in the simulation and recording the effect on the acceptance.

\subsection*{Tagged-photon flux \& target thickness}
The tagged-photon flux $N_\gamma$ was established by multiplying the FP hodoscope counts by the measured tagging efficiencies (${\sim}2$\% systematic uncertainty from tagging efficiency). The effective target thickness was
\begin{equation}
  \kappa_{\rm eff} = (8.14 \pm 0.10) \cdot 10^{23}\:\rm{nuclei/cm^2},
\end{equation}
with a ${\sim}1.2$\% systematic uncertainty originating from the geometry of the target. Further details about $N_\gamma$ and $\kappa_{\rm eff}$ can be found in Ref.~\cite{PhysRevC.98.012201}.

\begin{table}[b]
\begin{tabular}{ |c c R{1.8cm} C{1.8cm} | }
  \hline
  quantity           & source                  & magnitude          & in std.~dev. \\ \hline\hline
  $Y$                & fit                     & 2\%--7\%           & \checkmark   \\
                     & scattering              & ${\lesssim}5.5$\%  & \checkmark   \\
                     & Kapton                  & 1.5\%              &              \\ \hline
                                                                   
  $\rm \Omega_{eff}$ & positioning             & 3\%                & \checkmark   \\
                     & $E_{\rm cut}$           & 1.5\%              & \checkmark   \\
                     & model                   & 5\%                &              \\ \hline
                                                                   
  $N_\gamma$         & tagg.~eff.              & 2\%                & \checkmark   \\ \hline
                                                                   
  $\kappa_{\rm eff}$ & geometry                & 1.2\%              &              \\ \hline
                                                                   
  $P_{\rm c}$        & beam sim.               & ${\lesssim}1.6$\%  &              \\
                     & $\Delta E_{\gamma}$     & ${\lesssim}3.1$\%  &              \\ \hline
                                                                   
  $B\!R_{\rm rc}$    & measurement             & 1.5\%              &              \\ \hline

\end{tabular}
\caption{Summary of the dominant systematic uncertainties. The right column indicates if the systematic uncertainty contributes to the standard deviation of the nine cross-section measurements.}
\label{tab:sys_errs}
\end{table}

\subsection*{Pion-capture probability}
\begin{figure}[t]
    \centering
    \includegraphics[scale=0.45]{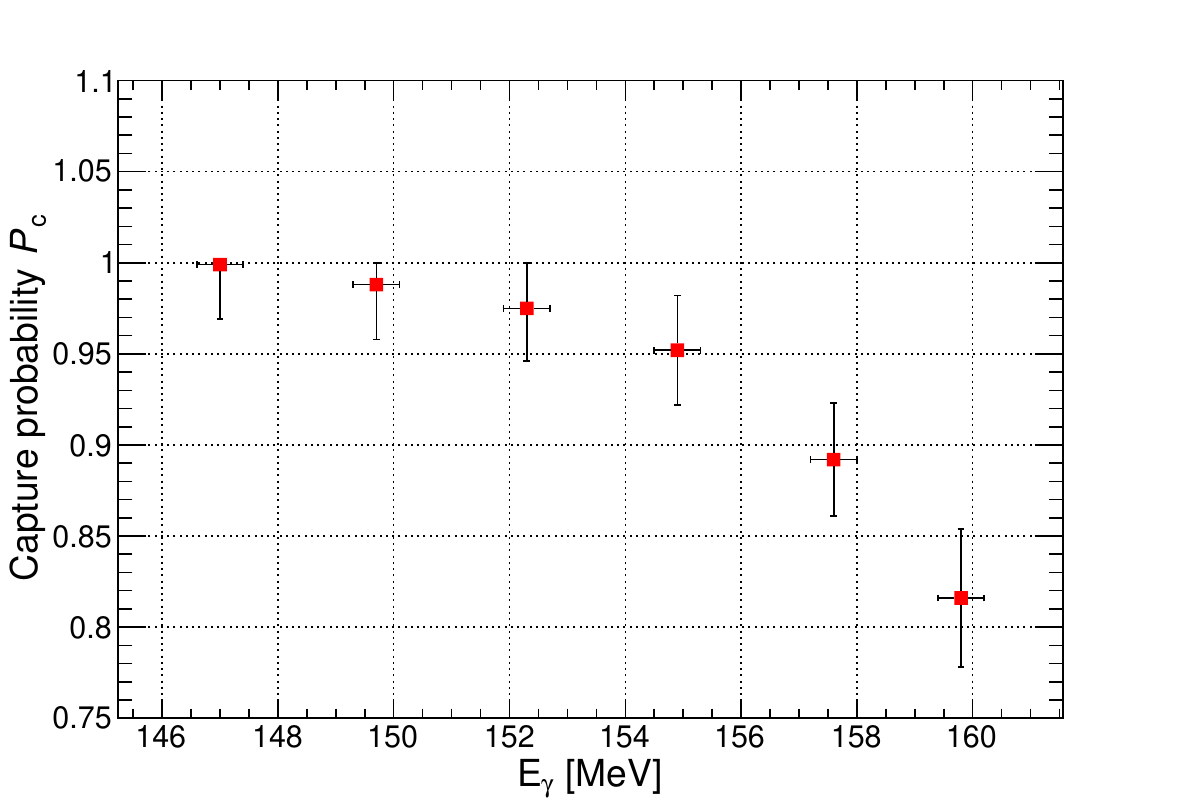}
    \caption{Probability of \PiM capture inside the LD$_2$ target with systematic uncertainties as a function of the incident photon energy $E_\gamma$.}
    \label{fig:capture}
\end{figure}

The capture probability of photoproduced \PiM $P_{\rm c}$ was estimated from a \geant\ simulation, where \PiM were simulated inside the LD$_2$ target. The X-Y coordinates of the vertices were sampled from a simulated intensity distribution of the photon beam determined by the geometry of the beam line, and the Z-coordinates (along the beam axis) were distributed uniformly over the length of the target. In sampling the momenta of the \PiM, the Fermi momentum of the bound neutron in the deuteron~\cite{LACOMBE1981139}, the energy of the incident photon and the angular distribution of the pions in the elementary photoproduction reaction~\cite{website:SAID} were taken into account. The dominant systematic uncertainty of ${\lesssim}$3.1\% originated from the ${\pm}0.4$~MeV uncertainty in the tagged-photon energies. The effect of uncertainty in the beam profile was estimated to be ${\lesssim}1.6$\% by changing the beam radius by $\pm10$\% in the simulation and recording the effect on $P_{\rm c}$. The simulated radius of the photon beam spot at the target center, $r_{\rm beam}\sim20$~mm, was in good agreement with a beam photograph at that location and was substantially smaller than the $r_{\rm vessel}=34$~mm radius of the Kapton vessel. Additionally, the \PiM escape from the target occurred predominantly from the downstream endcap, which explains the relatively weak dependence on the radius of the beam. Figure~\ref{fig:capture} depicts the dependence of $P_{\rm c}$ on the incident photon energy $E_\gamma$ with systematic uncertainties.

\subsection*{Results}
\begin{figure}[t]
  \centering
  \includegraphics[scale=0.45]{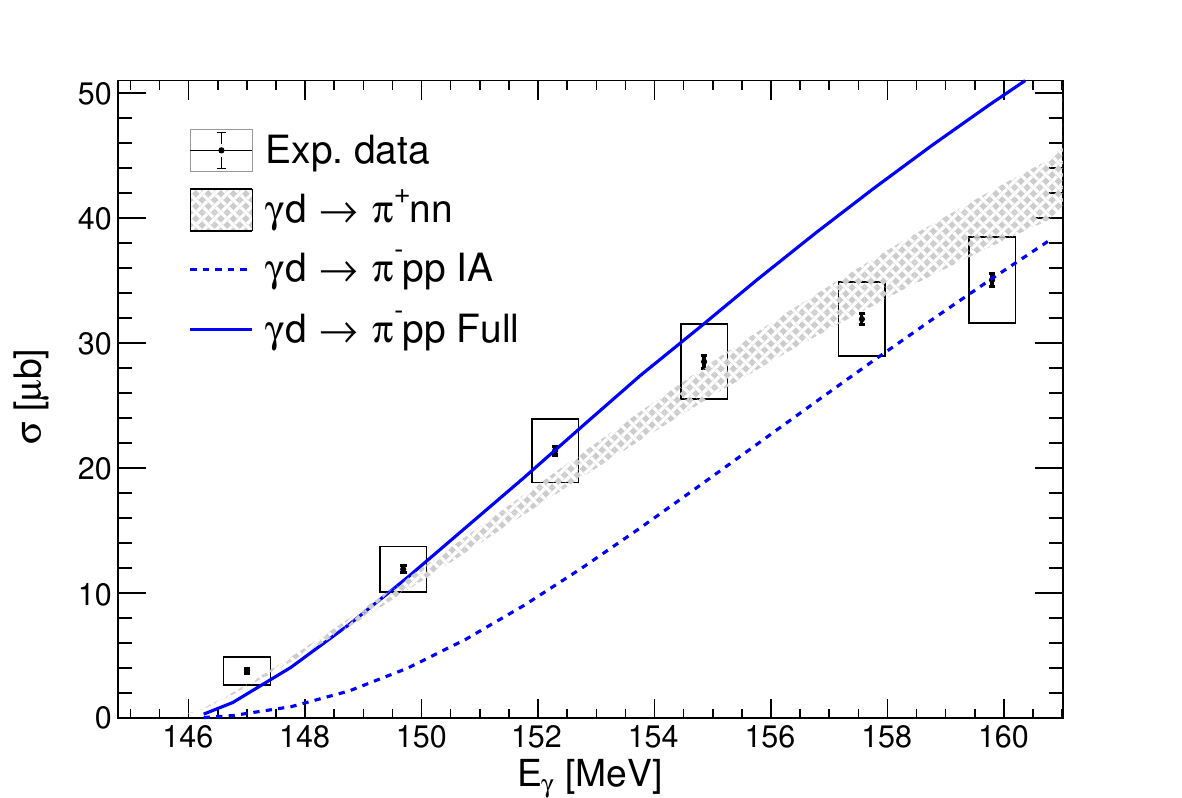}
  \caption{(Color online) Measured total cross section for \PiM photoproduction on the deuteron with statistical (error bars) and systematic (error boxes) uncertainties alongside theoretical predictions for $\gamma  d \rightarrow \pi^+ nn$ (gray band)~\cite{Lensky:2005hb} and $\gamma  d \rightarrow \pi^- pp$ in the Impulse Approximation (blue dashed line) and with FSI (blue solid line)~\cite{PhysRevC.84.035203}.}
  \label{fig:cs_plot}
\end{figure}

The cross section for threshold \PiM photoproduction at each energy was determined as a statistically weighted average of nine measurements (three detectors and three run periods). The standard deviation of the nine measurements was used to estimate the combined systematic uncertainty. Sources of systematic uncertainties are summarized in Table~\ref{tab:sys_errs}. The right column of Table~\ref{tab:sys_errs} specifies whether or not a given systematic uncertainty contributed to the standard deviation of the nine measurements. Typically, the uncertainty estimated from the standard deviation was of similar magnitude compared to the uncertainty estimated from adding the contributing sources in quadrature. The non-contributing sources were then added to the standard deviation in quadrature to produce the final systematic uncertainties (Table~\ref{tab:pim_xsec}, Fig.~\ref{fig:cs_plot}). Of the non-contributing uncertainties, only the capture efficiency affected the shape of the cross-section curve. Others affected the scale of the results. The angle- and energy-dependent uncertainties from scattering channels are accounted for in the standard deviation of the combined result, as they contributed to the observed spread of the nine measurements. A full account of the analysis of the experimental data is available in Ref.~\cite{Strandberg2017}.

\section{\label{sec:Theoretical-Analysis}Theoretical Analysis}

\begin{figure}[b]  
  \centering
  \includegraphics[scale=1]{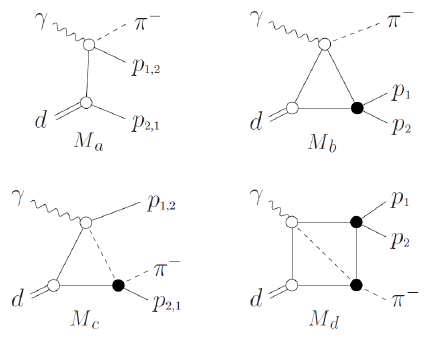}
  \vspace{-4mm}
  \caption{IA ($M_a$), $N\!N$-FSI ($M_b$), $\pi\!N$-FSI ($M_c$) and two-loop ($M_d$) diagram for $\gamma d \rightarrow\pi^- pp$. Filled black circles indicate FSI vertices.}
  \label{fig:diag}
\end{figure}

The experimental data for the $\gamma d\rightarrow\pi^{-}pp$ reaction are now compared with model predictions. Compared to the elementary reaction $\gamma n\rightarrow\pi^{-}p$, the additional final-state proton introduces additional FSI that have a non-negligible effect on the cross section. The model is a version of that described in Ref.~\cite{PhysRevC.84.035203}, simplified for the near-threshold region. It is calculated from the four diagrams displayed in Fig.~\ref{fig:diag}, where $M_{a}$ is the Impulse Approximation (IA) term, $M_{b}$ and $M_{c}$ are the $N\!N$ and $\pi\!N$ FSI terms, and $M_{d}$ is the $N\!N$-FSI term with pion rescattering in the intermediate state (the `two-loop' term). The ingredients and the approximations for the computation of the four terms are given in the following list.

\begin{table}[t]
\centering
\begin{tabular}{ | c | C{5cm} | }
  \hline
  $E_\gamma$~[MeV] & $\sigma \pm \: {\rm err_{stat.}} \pm \: {\rm err_{sys.}} $~[$\si{\micro\barn}$] \\ \hline \hline
  147.0 & $\phantom{0}3.8 \pm 0.2\:(5.3\%) \pm 1.1\:(28.9\%)$ \\ \hline
  149.7 & $11.9 \pm 0.3\:(2.5\%) \pm 1.8\:(15.1\%)$ \\ \hline
  152.3 & $21.4 \pm 0.3\:(1.4\%) \pm 2.5\:(11.7\%)$ \\ \hline
  154.9 & $28.5 \pm 0.5\:(1.8\%) \pm 3.0\:(10.5\%)$ \\ \hline
  157.6 & $31.9 \pm 0.4\:(1.3\%) \pm 3.0\:(\phantom{0}9.4\%)$ \\ \hline
  159.8 & $35.0 \pm 0.5\:(1.4\%) \pm 3.4\:(\phantom{0}9.7\%)$ \\ \hline 
\end{tabular}
\caption{Measured total cross section for \PiM photoproduction on the deuteron with statistical and systematic uncertainties.}
\label{tab:pim_xsec}
\end{table}

\begin{enumerate}
\item The elementary reaction is described by the $s$-wave amplitude, which is determined by the $E_{0+}$ multipole. The value of $E_{0+}$ as extracted by various analyses has been very stable over the last decades and here ${E_{0+}\!=-31.9}$ from Ref.~\cite{Drechsel:1992pn} is used. Here and elsewhere in the article the $E_{0+}$ amplitude is expressed in the conventional units of $10^{-3}/m_{\pi^+}$. Further, in diagrams $M_c$ and $M_d$ of Fig.~\ref{fig:diag}, only charged intermediate pions are included as the neutral-pion photoproduction amplitude is much smaller than the charged-pion photoproduction amplitude in the near-threshold region. In this approximation, the cross section is proportional to $\lvert E_{0+}\rvert ^2$.

\item The $s$-wave $pp$-scattering amplitude includes Coulomb effects and is taken in the Effective-Range Approximation~\cite{LandauBook}, using the values $a_{pp} = -7.8$~fm for the $pp$ scattering length and $r_{pp} = 2.8$~fm for the effective range. Off-shell effects are included as in Refs.~\cite{PhysRevC.74.014004,PhysRevC.84.035203}.

\item For $\pi\!N$ scattering, the $s$-wave $\pi\!N$ amplitude ${f_{\pi^- p} = b_0 - b_1}$ is used, fixed by the isospin scattering lengths $b_0 = -28$ and $b_1 = -881$ in units of $10^{-4}/m_{\pi}$~\cite{PhysRevC.70.045203}.

\item The deuteron wave function (DWF) derived from the Bonn potential is used in the parameterized form from Ref.~\cite{PhysRevC.63.024001}. The IA diagram $M_a$ includes both the $s$- and $d$-wave part of the DWF, with the $d$-wave having only a small effect on the cross section at energies close to threshold. The inclusion of the $d$-wave in other diagrams is expected to have a negligible effect and is neglected to simplify calculations.
\end{enumerate}

The cross-section model is compared with the experimental data in Fig.~\ref{fig:cs_plot}. The dashed curve indicates the IA ($M_a$ in Fig.~\ref{fig:diag}), whereas the solid curve indicates the full model (all terms in Fig.~\ref{fig:diag}). The dominant correction to the IA term $M_a$ originates from the $N\!N$-FSI amplitude $M_b$, whereas the combined contribution from the $\pi\!N$-FSI ($M_c$) and the two-loop term ($M_d$) is typically $\lesssim10$\%. Considering terms $M_c$ and $M_d$, the relative contribution of $M_c$ to the combined result of $M_a$ and $M_b$ is stable at around ${\sim}4$\%, while the effect of $M_d$ reduces from ${\sim}8$\% to ${\sim}2$\% as $E_\gamma$ increases from threshold to ${\sim}160$~MeV. While the model and the experimental data agree within uncertainties in the energy region 147~--~157~MeV, it overestimates the data above 157~MeV, due to two dominant factors:

\begin{enumerate}
\item The model does not account for the energy dependence of $E_{0+}$. Since $E_{0+}$ decreases with energy, this causes the theoretical model to overestimate the cross section as $E_{\gamma}$ increases from threshold.
\item The model uses only the $s$-wave amplitude for the elementary reaction $\gamma n\rightarrow\pi^{-}p$ and for $N\!N$-FSI, which is expected to contribute to the divergence as higher partial waves become significant at energies $\gtrsim10$~MeV above threshold.
\end{enumerate}

The measured cross section for $\gamma d \rightarrow \pi^-pp$ is also compared to a previous $\chi$PT prediction for the isospin-partner channel $\gamma d \rightarrow\pi^+nn$~\cite{Lensky:2005hb}. Comparison of the \PiM experimental data with the \PiP prediction is insightful as,  compared to Ref.~\cite{PhysRevC.84.035203}, the $\chi$PT calculation uses higher-order partial waves both for the elementary reaction ${\gamma p \rightarrow \pi^+ n}$ and for the $N\!N$-FSI. It also accounts for the energy dependence of $E_{0+}$. In the leading order of the chiral expansion, the elementary amplitudes $\gamma n \rightarrow \pi^- p$ and $\gamma p \rightarrow \pi^+ n$ are equal. The most important difference between the elementary \PiP and \PiM photoproduction reactions is the proton recoil in the latter, which increases the dipole moment of the final $\pi\!N$ system. Due to the absence of proton recoil in the \PiP reaction, the absolute value of $E_{0+}$ is approximately 12\% smaller for $\gamma p\rightarrow\pi^+n$ compared to $\gamma n\ \rightarrow\ \pi^- p$. For this calculation $E_{0+}(\pi^+n)=28.2$ from Ref.~\cite{BERNARD1996116} was used. This effect suppresses the cross section for $\gamma d \rightarrow\pi^+nn$ compared to $\gamma d \rightarrow \pi^-pp$. On the other hand, there is no Coulomb FSI in $\gamma d\rightarrow\pi^{+}nn$, which leads to a relative increase in the cross section compared to $\gamma d \rightarrow\pi^-pp$. These two effects are expected to cancel partially. The $\chi$PT calculation for $\gamma d \rightarrow\pi^+nn$ with theoretical uncertainties (see Ref.~\cite{Lensky:2005hb} for details of the uncertainty calculation) is depicted as a gray band in Fig.~\ref{fig:cs_plot}. The starting $E_\gamma$ value of the theoretical curve has been shifted to $145.8$~MeV to account for the difference in the reaction threshold compared to the \PiM channel. The calculation has been performed at the order ${\chi}^{5/2}$ of the chiral expansion parameter $\chi = m_\pi/m_N$, where $m_\pi$ ($m_N$) stands for the generic pion (nucleon) mass. The experimental data and the $\gamma d \rightarrow\pi^+nn$ model agree within uncertainties, suggesting that the differences between the \PiP and \PiM channels indeed tend largely to cancel. The good agreement between the models and the experimental data at energies $E_\gamma < 157$~MeV suggests that in the immediate vicinity of the threshold, the dominant processes that contribute to the cross section are relatively well understood.

\section{Summary}
In summary, the first measurement of the near-threshold cross section for \PiM photoproduction on the deuteron has been presented along with model predictions. The models and the experimental data are in good agreement in the vicinity of the threshold and provide new insight into the FSI behavior in this energy regime. The behavior further away from the threshold could be investigated by a dedicated $\chi$PT calculation for the measured $\gamma d\rightarrow\pi^{-}pp$ reaction. Further insight into the discrepancies between experimental data and the models at energies $\gtrsim10$~MeV above threshold could be gained from differential cross-section measurements for $\gamma d\rightarrow\pi^{-}pp$, which would allow for a more detailed study of the effects of various partial waves.

\begin{acknowledgments}
The authors would like to thank W.~R.~Gibbs, B.~F.~Gibson and G.~F.~de T\'eramond for constructive discussions. The authors acknowledge support from the staff of the MAX IV Laboratory. This research has been supported by the Swedish Research Council (Contracts No. 40324901 and No. 80410001), the Crafoord Foundation (Grant No. 20060749), the Scottish Universities Physics Alliance (SUPA), the SUPA Prize Studentship, UK STFC Grants No. 57071/1 and No. 50727/1, US NSF Grant No. PHY1309130, US DOE Grants No. \mbox{DE--SC0016581}, No. \mbox{DE--SC0016583} and No. \mbox{DE--FG02--06ER41422}, the NFS/IRES Award No. 0553467, RFBR Grants No. \mbox{16--02--00767} and \mbox{16--02--00767a}, and the Deutsche Forschungsgemeinschaft (DFG) through the Collaborative Research Center, SFB 1044.
\end{acknowledgments}

\bibliography{bstr_kf_etal}

\begin{thebibliography}{43}
\expandafter\ifx\csname natexlab\endcsname\relax\def\natexlab#1{#1}\fi
\expandafter\ifx\csname bibnamefont\endcsname\relax
  \def\bibnamefont#1{#1}\fi
\expandafter\ifx\csname bibfnamefont\endcsname\relax
  \def\bibfnamefont#1{#1}\fi
\expandafter\ifx\csname citenamefont\endcsname\relax
  \def\citenamefont#1{#1}\fi
\expandafter\ifx\csname url\endcsname\relax
  \def\url#1{\texttt{#1}}\fi
\expandafter\ifx\csname urlprefix\endcsname\relax\def\urlprefix{URL }\fi
\providecommand{\bibinfo}[2]{#2}
\providecommand{\eprint}[2][]{\url{#2}}

\bibitem[{\citenamefont{Drechsel and Tiator}(1992)}]{Drechsel:1992pn}
\bibinfo{author}{\bibfnamefont{D.}~\bibnamefont{Drechsel}} \bibnamefont{and}
  \bibinfo{author}{\bibfnamefont{L.}~\bibnamefont{Tiator}},
  \bibinfo{journal}{J. Phys. G} \textbf{\bibinfo{volume}{18}},
  \bibinfo{pages}{449} (\bibinfo{year}{1992}), \bibinfo{note}{and references
  therein}.

\bibitem[{\citenamefont{Arndt et~al.}(1990)\citenamefont{Arndt, Workman, Li,
  and Roper}}]{PhysRevC.42.1853}
\bibinfo{author}{\bibfnamefont{R.~A.} \bibnamefont{Arndt}},
  \bibinfo{author}{\bibfnamefont{R.~L.} \bibnamefont{Workman}},
  \bibinfo{author}{\bibfnamefont{Z.}~\bibnamefont{Li}}, \bibnamefont{and}
  \bibinfo{author}{\bibfnamefont{L.~D.} \bibnamefont{Roper}},
  \bibinfo{journal}{Phys. Rev. C} \textbf{\bibinfo{volume}{42}},
  \bibinfo{pages}{1853} (\bibinfo{year}{1990}).

\bibitem[{\citenamefont{Drechsel et~al.}(2007)\citenamefont{Drechsel, Pasquini,
  and Tiator}}]{Drechsel2007}
\bibinfo{author}{\bibfnamefont{D.}~\bibnamefont{Drechsel}},
  \bibinfo{author}{\bibfnamefont{B.}~\bibnamefont{Pasquini}}, \bibnamefont{and}
  \bibinfo{author}{\bibfnamefont{L.}~\bibnamefont{Tiator}},
  \bibinfo{journal}{Few-Body Syst.} \textbf{\bibinfo{volume}{41}},
  \bibinfo{pages}{13} (\bibinfo{year}{2007}).

\bibitem[{\citenamefont{Hilt et~al.}(2013)\citenamefont{Hilt, Lehnhart,
  Scherer, and Tiator}}]{PhysRevC.88.055207}
\bibinfo{author}{\bibfnamefont{M.}~\bibnamefont{Hilt}},
  \bibinfo{author}{\bibfnamefont{B.~C.} \bibnamefont{Lehnhart}},
  \bibinfo{author}{\bibfnamefont{S.}~\bibnamefont{Scherer}}, \bibnamefont{and}
  \bibinfo{author}{\bibfnamefont{L.}~\bibnamefont{Tiator}},
  \bibinfo{journal}{Phys. Rev. C} \textbf{\bibinfo{volume}{88}},
  \bibinfo{pages}{055207} (\bibinfo{year}{2013}), \bibinfo{note}{and references
  therein}.

\bibitem[{\citenamefont{Bergstrom et~al.}(1996)\citenamefont{Bergstrom, Vogt,
  Igarashi, Keeter, Hallin, Retzlaff, Skopik, and Booth}}]{PhysRevC.53.R1052}
\bibinfo{author}{\bibfnamefont{J.~C.} \bibnamefont{Bergstrom}},
  \bibinfo{author}{\bibfnamefont{J.~M.} \bibnamefont{Vogt}},
  \bibinfo{author}{\bibfnamefont{R.}~\bibnamefont{Igarashi}},
  \bibinfo{author}{\bibfnamefont{K.~J.} \bibnamefont{Keeter}},
  \bibinfo{author}{\bibfnamefont{E.~L.} \bibnamefont{Hallin}},
  \bibinfo{author}{\bibfnamefont{G.~A.} \bibnamefont{Retzlaff}},
  \bibinfo{author}{\bibfnamefont{D.~M.} \bibnamefont{Skopik}},
  \bibnamefont{and} \bibinfo{author}{\bibfnamefont{E.~C.} \bibnamefont{Booth}},
  \bibinfo{journal}{Phys. Rev. C} \textbf{\bibinfo{volume}{53}},
  \bibinfo{pages}{1052} (\bibinfo{year}{1996}).

\bibitem[{\citenamefont{Bergstrom et~al.}(1997)\citenamefont{Bergstrom,
  Igarashi, and Vogt}}]{PhysRevC.55.2016}
\bibinfo{author}{\bibfnamefont{J.~C.} \bibnamefont{Bergstrom}},
  \bibinfo{author}{\bibfnamefont{R.}~\bibnamefont{Igarashi}}, \bibnamefont{and}
  \bibinfo{author}{\bibfnamefont{J.~M.} \bibnamefont{Vogt}},
  \bibinfo{journal}{Phys. Rev. C} \textbf{\bibinfo{volume}{55}},
  \bibinfo{pages}{2016} (\bibinfo{year}{1997}).

\bibitem[{\citenamefont{Bergstrom}(1998)}]{PhysRevC.58.2574}
\bibinfo{author}{\bibfnamefont{J.~C.} \bibnamefont{Bergstrom}},
  \bibinfo{journal}{Phys. Rev. C} \textbf{\bibinfo{volume}{58}},
  \bibinfo{pages}{2574} (\bibinfo{year}{1998}).

\bibitem[{\citenamefont{Hornidge et~al.}(2013)\citenamefont{Hornidge,
  Aguar~Bartolom\'e, Annand, Arends, Beck, Bekrenev, Bergh\"auser, Bernstein,
  Braghieri, Briscoe et~al.}}]{PhysRevLett.111.062004}
\bibinfo{author}{\bibfnamefont{D.}~\bibnamefont{Hornidge}},
  \bibinfo{author}{\bibfnamefont{P.}~\bibnamefont{Aguar~Bartolom\'e}},
  \bibinfo{author}{\bibfnamefont{J.~R.~M.} \bibnamefont{Annand}},
  \bibinfo{author}{\bibfnamefont{H.~J.} \bibnamefont{Arends}},
  \bibinfo{author}{\bibfnamefont{R.}~\bibnamefont{Beck}},
  \bibinfo{author}{\bibfnamefont{V.}~\bibnamefont{Bekrenev}},
  \bibinfo{author}{\bibfnamefont{H.}~\bibnamefont{Bergh\"auser}},
  \bibinfo{author}{\bibfnamefont{A.~M.} \bibnamefont{Bernstein}},
  \bibinfo{author}{\bibfnamefont{A.}~\bibnamefont{Braghieri}},
  \bibinfo{author}{\bibfnamefont{W.~J.} \bibnamefont{Briscoe}},
  \bibnamefont{et~al.} (\bibinfo{collaboration}{A2 Collaboration and CB-TAPS
  Collaboration}), \bibinfo{journal}{Phys. Rev. Lett.}
  \textbf{\bibinfo{volume}{111}}, \bibinfo{pages}{062004}
  (\bibinfo{year}{2013}).

\bibitem[{\citenamefont{Booth et~al.}(1979)\citenamefont{Booth, Chasan,
  Comuzzi, and Bosted}}]{Booth_1979}
\bibinfo{author}{\bibfnamefont{E.~C.} \bibnamefont{Booth}},
  \bibinfo{author}{\bibfnamefont{B.}~\bibnamefont{Chasan}},
  \bibinfo{author}{\bibfnamefont{J.}~\bibnamefont{Comuzzi}}, \bibnamefont{and}
  \bibinfo{author}{\bibfnamefont{P.}~\bibnamefont{Bosted}},
  \bibinfo{journal}{Phys. Rev. C} \textbf{\bibinfo{volume}{20}},
  \bibinfo{pages}{1217} (\bibinfo{year}{1979}).

\bibitem[{\citenamefont{Rossi et~al.}(1973)\citenamefont{Rossi, Piazza,
  Susinno, Carbonara, Gialanella, Napolitano, Rinzivillo, Votano, Mantovani,
  Piazzoli et~al.}}]{Rossi1973}
\bibinfo{author}{\bibfnamefont{V.}~\bibnamefont{Rossi}},
  \bibinfo{author}{\bibfnamefont{A.}~\bibnamefont{Piazza}},
  \bibinfo{author}{\bibfnamefont{G.}~\bibnamefont{Susinno}},
  \bibinfo{author}{\bibfnamefont{F.}~\bibnamefont{Carbonara}},
  \bibinfo{author}{\bibfnamefont{G.}~\bibnamefont{Gialanella}},
  \bibinfo{author}{\bibfnamefont{M.}~\bibnamefont{Napolitano}},
  \bibinfo{author}{\bibfnamefont{R.}~\bibnamefont{Rinzivillo}},
  \bibinfo{author}{\bibfnamefont{L.}~\bibnamefont{Votano}},
  \bibinfo{author}{\bibfnamefont{G.~C.} \bibnamefont{Mantovani}},
  \bibinfo{author}{\bibfnamefont{A.}~\bibnamefont{Piazzoli}},
  \bibnamefont{et~al.}, \bibinfo{journal}{Nuovo Cimento A}
  \textbf{\bibinfo{volume}{13}}, \bibinfo{pages}{59} (\bibinfo{year}{1973}).

\bibitem[{\citenamefont{Salomon et~al.}(1984)\citenamefont{Salomon, Measday,
  Poutissou, and Robertson}}]{SALOMON1984493}
\bibinfo{author}{\bibfnamefont{M.}~\bibnamefont{Salomon}},
  \bibinfo{author}{\bibfnamefont{D.}~\bibnamefont{Measday}},
  \bibinfo{author}{\bibfnamefont{J.-M.} \bibnamefont{Poutissou}},
  \bibnamefont{and}
  \bibinfo{author}{\bibfnamefont{B.}~\bibnamefont{Robertson}},
  \bibinfo{journal}{Nucl. Phys. A} \textbf{\bibinfo{volume}{414}},
  \bibinfo{pages}{493 } (\bibinfo{year}{1984}), ISSN \bibinfo{issn}{0375-9474}.

\bibitem[{\citenamefont{Wang}(1992)}]{Wang1992}
\bibinfo{author}{\bibfnamefont{M.}~\bibnamefont{Wang}}, Ph.D. thesis,
  \bibinfo{school}{University Of Kentucky} (\bibinfo{year}{1992}).

\bibitem[{\citenamefont{Liu}(1994)}]{Liu1994}
\bibinfo{author}{\bibfnamefont{K.}~\bibnamefont{Liu}}, Ph.D. thesis,
  \bibinfo{school}{University Of Kentucky} (\bibinfo{year}{1994}).

\bibitem[{\citenamefont{Gabioud et~al.}(1979)\citenamefont{Gabioud,
  \mbox{J.-C.} Alder, Joseph, \mbox{J.-F.} Loude, Morel, Perrenoud,
  \mbox{J.-P.} Perroud, Tran, Winkelmann, Dahme et~al.}}]{gabioud1979}
\bibinfo{author}{\bibfnamefont{B.}~\bibnamefont{Gabioud}},
  \bibinfo{author}{\bibnamefont{\mbox{J.-C.} Alder}},
  \bibinfo{author}{\bibfnamefont{C.}~\bibnamefont{Joseph}},
  \bibinfo{author}{\bibnamefont{\mbox{J.-F.} Loude}},
  \bibinfo{author}{\bibfnamefont{N.}~\bibnamefont{Morel}},
  \bibinfo{author}{\bibfnamefont{A.}~\bibnamefont{Perrenoud}},
  \bibinfo{author}{\bibnamefont{\mbox{J.-P.} Perroud}},
  \bibinfo{author}{\bibfnamefont{M.~T.} \bibnamefont{Tran}},
  \bibinfo{author}{\bibfnamefont{E.}~\bibnamefont{Winkelmann}},
  \bibinfo{author}{\bibfnamefont{W.}~\bibnamefont{Dahme}},
  \bibnamefont{et~al.}, \bibinfo{journal}{Phys.\ Rev.\ Lett.}
  \textbf{\bibinfo{volume}{42}}, \bibinfo{pages}{1508} (\bibinfo{year}{1979}).

\bibitem[{\citenamefont{\mbox{J.-O.} Adler
  et~al.}(2013)\citenamefont{\mbox{J.-O.} Adler, Boland, Brudvik, Fissum,
  Hansen, Isaksson, Lilja, \mbox{L.-J.} Lindgren, Lundin, Nilsson
  et~al.}}]{adler2012}
\bibinfo{author}{\bibnamefont{\mbox{J.-O.} Adler}},
  \bibinfo{author}{\bibfnamefont{M.}~\bibnamefont{Boland}},
  \bibinfo{author}{\bibfnamefont{J.}~\bibnamefont{Brudvik}},
  \bibinfo{author}{\bibfnamefont{K.}~\bibnamefont{Fissum}},
  \bibinfo{author}{\bibfnamefont{K.}~\bibnamefont{Hansen}},
  \bibinfo{author}{\bibfnamefont{L.}~\bibnamefont{Isaksson}},
  \bibinfo{author}{\bibfnamefont{P.}~\bibnamefont{Lilja}},
  \bibinfo{author}{\bibnamefont{\mbox{L.-J.} Lindgren}},
  \bibinfo{author}{\bibfnamefont{M.}~\bibnamefont{Lundin}},
  \bibinfo{author}{\bibfnamefont{B.}~\bibnamefont{Nilsson}},
  \bibnamefont{et~al.}, \bibinfo{journal}{Nucl.\ Instrum.\ Methods\ Phys.\
  Res.\ Sect.\ A} \textbf{\bibinfo{volume}{715}}, \bibinfo{pages}{1}
  (\bibinfo{year}{2013}).

\bibitem[{\citenamefont{Eriksson}(2014)}]{eriksson2014}
\bibinfo{author}{\bibfnamefont{M.}~\bibnamefont{Eriksson}}, in
  \emph{\bibinfo{booktitle}{Proceedings of IPAC, Dresden, Germany}}
  (\bibinfo{year}{2014}).

\bibitem[{\citenamefont{\mbox{J.-O.} Adler
  et~al.}(1990)\citenamefont{\mbox{J.-O.} Adler, \mbox{B.-E.} Andersson,
  \mbox{K.~I.} Blomqvist, Forkman, Hansen, Isaksson, Lindgren, Nilsson,
  Sandell, Schr{\"o}der et~al.}}]{adler1990}
\bibinfo{author}{\bibnamefont{\mbox{J.-O.} Adler}},
  \bibinfo{author}{\bibnamefont{\mbox{B.-E.} Andersson}},
  \bibinfo{author}{\bibnamefont{\mbox{K.~I.} Blomqvist}},
  \bibinfo{author}{\bibfnamefont{B.}~\bibnamefont{Forkman}},
  \bibinfo{author}{\bibfnamefont{K.}~\bibnamefont{Hansen}},
  \bibinfo{author}{\bibfnamefont{L.}~\bibnamefont{Isaksson}},
  \bibinfo{author}{\bibfnamefont{K.}~\bibnamefont{Lindgren}},
  \bibinfo{author}{\bibfnamefont{D.}~\bibnamefont{Nilsson}},
  \bibinfo{author}{\bibfnamefont{A.}~\bibnamefont{Sandell}},
  \bibinfo{author}{\bibfnamefont{B.}~\bibnamefont{Schr{\"o}der}},
  \bibnamefont{et~al.}, \bibinfo{journal}{Nucl.\ Instrum.\ Methods\ Phys.\
  Res.\ Sect.\ A} \textbf{\bibinfo{volume}{294}}, \bibinfo{pages}{15}
  (\bibinfo{year}{1990}).

\bibitem[{\citenamefont{\mbox{J.-O.} Adler
  et~al.}(1997)\citenamefont{\mbox{J.-O.} Adler, \mbox{B.-E.} Andersson,
  \mbox{K.~I.} Blomqvist, \mbox{K.~G.} Fissum, Hansen, Isaksson, Nilsson,
  Nilsson, Ruijter, Sandell et~al.}}]{adler1997}
\bibinfo{author}{\bibnamefont{\mbox{J.-O.} Adler}},
  \bibinfo{author}{\bibnamefont{\mbox{B.-E.} Andersson}},
  \bibinfo{author}{\bibnamefont{\mbox{K.~I.} Blomqvist}},
  \bibinfo{author}{\bibnamefont{\mbox{K.~G.} Fissum}},
  \bibinfo{author}{\bibfnamefont{K.}~\bibnamefont{Hansen}},
  \bibinfo{author}{\bibfnamefont{L.}~\bibnamefont{Isaksson}},
  \bibinfo{author}{\bibfnamefont{B.}~\bibnamefont{Nilsson}},
  \bibinfo{author}{\bibfnamefont{D.}~\bibnamefont{Nilsson}},
  \bibinfo{author}{\bibfnamefont{H.}~\bibnamefont{Ruijter}},
  \bibinfo{author}{\bibfnamefont{A.}~\bibnamefont{Sandell}},
  \bibnamefont{et~al.}, \bibinfo{journal}{Nucl.\ Instrum.\ Methods\ Phys.\
  Res.\ Sect.\ A} \textbf{\bibinfo{volume}{388}}, \bibinfo{pages}{17}
  (\bibinfo{year}{1997}).

\bibitem[{\citenamefont{\mbox{J. M.} Vogt et~al.}(1993)\citenamefont{\mbox{J.
  M.} Vogt, \mbox{R. E.} Pywell, \mbox{D. M.} Skopik, \mbox{E. L.} Hallin,
  \mbox{J. C.} Bergstrom, \mbox{H. S.} Caplan, \mbox{K. I.} Blomqvist, Bianco,
  and \mbox{J. W.} Jury}}]{vogt1993}
\bibinfo{author}{\bibnamefont{\mbox{J. M.} Vogt}},
  \bibinfo{author}{\bibnamefont{\mbox{R. E.} Pywell}},
  \bibinfo{author}{\bibnamefont{\mbox{D. M.} Skopik}},
  \bibinfo{author}{\bibnamefont{\mbox{E. L.} Hallin}},
  \bibinfo{author}{\bibnamefont{\mbox{J. C.} Bergstrom}},
  \bibinfo{author}{\bibnamefont{\mbox{H. S.} Caplan}},
  \bibinfo{author}{\bibnamefont{\mbox{K. I.} Blomqvist}},
  \bibinfo{author}{\bibfnamefont{W.~D.} \bibnamefont{Bianco}},
  \bibnamefont{and} \bibinfo{author}{\bibnamefont{\mbox{J. W.} Jury}},
  \bibinfo{journal}{Nucl.\ Instrum.\ Methods\ Phys.\ Res.\ Sect.\ A}
  \textbf{\bibinfo{volume}{324}}, \bibinfo{pages}{198} (\bibinfo{year}{1993}).

\bibitem[{\citenamefont{\mbox{J. P.} Miller et~al.}(1988)\citenamefont{\mbox{J.
  P.} Miller, \mbox{E. J.} Austin, \mbox{E. C.} Booth, \mbox{K. P.} Gall,
  \mbox{E. K.} McIntyre, and \mbox{D. A.} Whitehouse}}]{miller1988}
\bibinfo{author}{\bibnamefont{\mbox{J. P.} Miller}},
  \bibinfo{author}{\bibnamefont{\mbox{E. J.} Austin}},
  \bibinfo{author}{\bibnamefont{\mbox{E. C.} Booth}},
  \bibinfo{author}{\bibnamefont{\mbox{K. P.} Gall}},
  \bibinfo{author}{\bibnamefont{\mbox{E. K.} McIntyre}}, \bibnamefont{and}
  \bibinfo{author}{\bibnamefont{\mbox{D. A.} Whitehouse}},
  \bibinfo{journal}{Nucl.\ Instrum.\ Methods\ Phys.\ Res.\ Sect.\ A}
  \textbf{\bibinfo{volume}{270}}, \bibinfo{pages}{431} (\bibinfo{year}{1988}).

\bibitem[{\citenamefont{H{\"u}nger et~al.}(1997)\citenamefont{H{\"u}nger,
  Peise, Robbiano, Ahrens, Anthony, Arends, Beck, Capitani, Dolbilkin,
  Falkenberg et~al.}}]{Hunger1997385}
\bibinfo{author}{\bibfnamefont{A.}~\bibnamefont{H{\"u}nger}},
  \bibinfo{author}{\bibfnamefont{J.}~\bibnamefont{Peise}},
  \bibinfo{author}{\bibfnamefont{A.}~\bibnamefont{Robbiano}},
  \bibinfo{author}{\bibfnamefont{J.}~\bibnamefont{Ahrens}},
  \bibinfo{author}{\bibfnamefont{I.}~\bibnamefont{Anthony}},
  \bibinfo{author}{\bibfnamefont{H.-J.} \bibnamefont{Arends}},
  \bibinfo{author}{\bibfnamefont{R.}~\bibnamefont{Beck}},
  \bibinfo{author}{\bibfnamefont{G.}~\bibnamefont{Capitani}},
  \bibinfo{author}{\bibfnamefont{B.}~\bibnamefont{Dolbilkin}},
  \bibinfo{author}{\bibfnamefont{H.}~\bibnamefont{Falkenberg}},
  \bibnamefont{et~al.}, \bibinfo{journal}{Nucl. Phys. A}
  \textbf{\bibinfo{volume}{620}}, \bibinfo{pages}{385 } (\bibinfo{year}{1997}).

\bibitem[{\citenamefont{\mbox{L. S.} Myers}(2010)}]{myers2010}
\bibinfo{author}{\bibnamefont{\mbox{L. S.} Myers}}, Ph.D. thesis,
  \bibinfo{school}{University of Illinois at Urbana-Champaign, Champaign, IL,
  USA} (\bibinfo{year}{2010}).

\bibitem[{\citenamefont{Brun and Rademakers}(1997)}]{Brun199781}
\bibinfo{author}{\bibfnamefont{R.}~\bibnamefont{Brun}} \bibnamefont{and}
  \bibinfo{author}{\bibfnamefont{F.}~\bibnamefont{Rademakers}},
  \bibinfo{journal}{Nucl. Instrum. Meth. Phys. Res. A}
  \textbf{\bibinfo{volume}{389}}, \bibinfo{pages}{81 } (\bibinfo{year}{1997}).

\bibitem[{\citenamefont{Verkerke and Kirkby}(2003)}]{Verkerke:2003ir}
\bibinfo{author}{\bibfnamefont{W.}~\bibnamefont{Verkerke}} \bibnamefont{and}
  \bibinfo{author}{\bibfnamefont{D.}~\bibnamefont{Kirkby}},
  \bibinfo{journal}{eConf} \textbf{\bibinfo{volume}{C0303241}}
  (\bibinfo{year}{2003}).

\bibitem[{\citenamefont{Strandberg et~al.}(2018)\citenamefont{Strandberg,
  Annand, Briscoe, Brudvik, Cividini, Clark, Downie, England, Feldman, Fissum
  et~al.}}]{PhysRevC.98.012201}
\bibinfo{author}{\bibfnamefont{B.}~\bibnamefont{Strandberg}},
  \bibinfo{author}{\bibfnamefont{J.~R.~M.} \bibnamefont{Annand}},
  \bibinfo{author}{\bibfnamefont{W.}~\bibnamefont{Briscoe}},
  \bibinfo{author}{\bibfnamefont{J.}~\bibnamefont{Brudvik}},
  \bibinfo{author}{\bibfnamefont{F.}~\bibnamefont{Cividini}},
  \bibinfo{author}{\bibfnamefont{L.}~\bibnamefont{Clark}},
  \bibinfo{author}{\bibfnamefont{E.~J.} \bibnamefont{Downie}},
  \bibinfo{author}{\bibfnamefont{K.}~\bibnamefont{England}},
  \bibinfo{author}{\bibfnamefont{G.}~\bibnamefont{Feldman}},
  \bibinfo{author}{\bibfnamefont{K.~G.} \bibnamefont{Fissum}},
  \bibnamefont{et~al.} (\bibinfo{collaboration}{The COMPTON@MAX-lab
  Collaboration}), \bibinfo{journal}{Phys. Rev. C}
  \textbf{\bibinfo{volume}{98}}, \bibinfo{pages}{012201}
  (\bibinfo{year}{2018}).

\bibitem[{\citenamefont{Highland et~al.}(1981)\citenamefont{Highland, Salomon,
  Hasinoff, Mazzucato, Measday, Poutissou, and Suzuki}}]{Highland1981333}
\bibinfo{author}{\bibfnamefont{V.~L.} \bibnamefont{Highland}},
  \bibinfo{author}{\bibfnamefont{M.}~\bibnamefont{Salomon}},
  \bibinfo{author}{\bibfnamefont{M.}~\bibnamefont{Hasinoff}},
  \bibinfo{author}{\bibfnamefont{E.}~\bibnamefont{Mazzucato}},
  \bibinfo{author}{\bibfnamefont{D.}~\bibnamefont{Measday}},
  \bibinfo{author}{\bibfnamefont{J.-M.} \bibnamefont{Poutissou}},
  \bibnamefont{and} \bibinfo{author}{\bibfnamefont{T.}~\bibnamefont{Suzuki}},
  \bibinfo{journal}{Nucl. Phys. A} \textbf{\bibinfo{volume}{365}},
  \bibinfo{pages}{333 } (\bibinfo{year}{1981}).

\bibitem[{\citenamefont{Gibbs et~al.}(1975)\citenamefont{Gibbs, Gibson, and
  {Stephenson, Jr.}}}]{PhysRevC.11.90}
\bibinfo{author}{\bibfnamefont{W.~R.} \bibnamefont{Gibbs}},
  \bibinfo{author}{\bibfnamefont{B.~F.} \bibnamefont{Gibson}},
  \bibnamefont{and} \bibinfo{author}{\bibfnamefont{G.~J.}
  \bibnamefont{{Stephenson, Jr.}}}, \bibinfo{journal}{Phys. Rev. C}
  \textbf{\bibinfo{volume}{11}}, \bibinfo{pages}{90} (\bibinfo{year}{1975}).

\bibitem[{\citenamefont{de~T\'eramond}(1977)}]{PhysRevC.16.1976}
\bibinfo{author}{\bibfnamefont{G.~F.} \bibnamefont{de~T\'eramond}},
  \bibinfo{journal}{Phys. Rev. C} \textbf{\bibinfo{volume}{16}},
  \bibinfo{pages}{1976} (\bibinfo{year}{1977}).

\bibitem[{\citenamefont{Gibbs and Gibson}()}]{GibbsGibson}
\bibinfo{author}{\bibfnamefont{W.~R.} \bibnamefont{Gibbs}} \bibnamefont{and}
  \bibinfo{author}{\bibfnamefont{B.~F.} \bibnamefont{Gibson}},
  \bibinfo{note}{private communication}.

\bibitem[{\citenamefont{Allison et~al.}(2016)\citenamefont{Allison, Amako,
  Apostolakis, Arce, Asai, Aso, Bagli, Bagulya, Banerjee, Barrand
  et~al.}}]{Geant4}
\bibinfo{author}{\bibfnamefont{J.}~\bibnamefont{Allison}},
  \bibinfo{author}{\bibfnamefont{K.}~\bibnamefont{Amako}},
  \bibinfo{author}{\bibfnamefont{J.}~\bibnamefont{Apostolakis}},
  \bibinfo{author}{\bibfnamefont{P.}~\bibnamefont{Arce}},
  \bibinfo{author}{\bibfnamefont{M.}~\bibnamefont{Asai}},
  \bibinfo{author}{\bibfnamefont{T.}~\bibnamefont{Aso}},
  \bibinfo{author}{\bibfnamefont{E.}~\bibnamefont{Bagli}},
  \bibinfo{author}{\bibfnamefont{A.}~\bibnamefont{Bagulya}},
  \bibinfo{author}{\bibfnamefont{S.}~\bibnamefont{Banerjee}},
  \bibinfo{author}{\bibfnamefont{G.}~\bibnamefont{Barrand}},
  \bibnamefont{et~al.}, \bibinfo{journal}{Nucl. Instrum. Meth. Phys. Res. A}
  \textbf{\bibinfo{volume}{835}}, \bibinfo{pages}{186 } (\bibinfo{year}{2016}).

\bibitem[{\citenamefont{Myers et~al.}(2015)\citenamefont{Myers, Annand,
  Brudvik, Feldman, Fissum, Grie\ss{}hammer, Hansen, Henshaw, Isaksson, Jebali
  et~al.}}]{PhysRevC.92.025203}
\bibinfo{author}{\bibfnamefont{L.~S.} \bibnamefont{Myers}},
  \bibinfo{author}{\bibfnamefont{J.~R.~M.} \bibnamefont{Annand}},
  \bibinfo{author}{\bibfnamefont{J.}~\bibnamefont{Brudvik}},
  \bibinfo{author}{\bibfnamefont{G.}~\bibnamefont{Feldman}},
  \bibinfo{author}{\bibfnamefont{K.~G.} \bibnamefont{Fissum}},
  \bibinfo{author}{\bibfnamefont{H.~W.} \bibnamefont{Grie\ss{}hammer}},
  \bibinfo{author}{\bibfnamefont{K.}~\bibnamefont{Hansen}},
  \bibinfo{author}{\bibfnamefont{S.~S.} \bibnamefont{Henshaw}},
  \bibinfo{author}{\bibfnamefont{L.}~\bibnamefont{Isaksson}},
  \bibinfo{author}{\bibfnamefont{R.}~\bibnamefont{Jebali}},
  \bibnamefont{et~al.}, \bibinfo{journal}{Phys. Rev. C}
  \textbf{\bibinfo{volume}{92}}, \bibinfo{pages}{025203}
  (\bibinfo{year}{2015}).

\bibitem[{\citenamefont{Kossert et~al.}(2002)\citenamefont{Kossert, Camen,
  Wissmann, Ahrens, Annand, Arends, Beck, Caselotti, Grabmayr, Jahn
  et~al.}}]{PhysRevLett.88.162301}
\bibinfo{author}{\bibfnamefont{K.}~\bibnamefont{Kossert}},
  \bibinfo{author}{\bibfnamefont{M.}~\bibnamefont{Camen}},
  \bibinfo{author}{\bibfnamefont{F.}~\bibnamefont{Wissmann}},
  \bibinfo{author}{\bibfnamefont{J.}~\bibnamefont{Ahrens}},
  \bibinfo{author}{\bibfnamefont{J.~R.~M.} \bibnamefont{Annand}},
  \bibinfo{author}{\bibfnamefont{H.-J.} \bibnamefont{Arends}},
  \bibinfo{author}{\bibfnamefont{R.}~\bibnamefont{Beck}},
  \bibinfo{author}{\bibfnamefont{G.}~\bibnamefont{Caselotti}},
  \bibinfo{author}{\bibfnamefont{P.}~\bibnamefont{Grabmayr}},
  \bibinfo{author}{\bibfnamefont{O.}~\bibnamefont{Jahn}}, \bibnamefont{et~al.},
  \bibinfo{journal}{Phys. Rev. Lett.} \textbf{\bibinfo{volume}{88}},
  \bibinfo{pages}{162301} (\bibinfo{year}{2002}).

\bibitem[{\citenamefont{\mbox{R.} Al~Jebali}(2013)}]{jebali2013}
\bibinfo{author}{\bibnamefont{\mbox{R.} Al~Jebali}}, Ph.D. thesis,
  \bibinfo{school}{The University Of Glasgow, Glasgow, UK}
  (\bibinfo{year}{2013}).

\bibitem[{\citenamefont{Lacombe et~al.}(1981)\citenamefont{Lacombe, Loiseau,
  Mau, C\^ot\'e, Pir\'es, and de~Tourreil}}]{LACOMBE1981139}
\bibinfo{author}{\bibfnamefont{M.}~\bibnamefont{Lacombe}},
  \bibinfo{author}{\bibfnamefont{B.}~\bibnamefont{Loiseau}},
  \bibinfo{author}{\bibfnamefont{R.}~\bibnamefont{Mau}},
  \bibinfo{author}{\bibfnamefont{J.}~\bibnamefont{C\^ot\'e}},
  \bibinfo{author}{\bibfnamefont{P.}~\bibnamefont{Pir\'es}}, \bibnamefont{and}
  \bibinfo{author}{\bibfnamefont{R.}~\bibnamefont{de~Tourreil}},
  \bibinfo{journal}{Phys. Lett. B} \textbf{\bibinfo{volume}{101}},
  \bibinfo{pages}{139} (\bibinfo{year}{1981}).

\bibitem[{web()}]{website:SAID}
\bibinfo{note}{W. J. Briscoe, I. I. Strakovsky, and R. L. Workman, Institute of
  Nuclear Studies of The George Washington University Database [SAID];
  \url{http://gwdac.phys.gwu.edu}}.

\bibitem[{\citenamefont{\mbox{B} Strandberg}(2017)}]{Strandberg2017}
\bibinfo{author}{\bibnamefont{\mbox{B} Strandberg}}, Ph.D. thesis,
  \bibinfo{school}{University Of Glasgow, Glasgow, UK} (\bibinfo{year}{2017}).

\bibitem[{\citenamefont{Lensky et~al.}(2005)\citenamefont{Lensky, Baru,
  Haidenbauer, Hanhart, Kudryavtsev, and Mei{\ss}ner}}]{Lensky:2005hb}
\bibinfo{author}{\bibfnamefont{V.}~\bibnamefont{Lensky}},
  \bibinfo{author}{\bibfnamefont{V.}~\bibnamefont{Baru}},
  \bibinfo{author}{\bibfnamefont{J.}~\bibnamefont{Haidenbauer}},
  \bibinfo{author}{\bibfnamefont{C.}~\bibnamefont{Hanhart}},
  \bibinfo{author}{\bibfnamefont{A.}~\bibnamefont{Kudryavtsev}},
  \bibnamefont{and} \bibinfo{author}{\bibfnamefont{U.~G.}
  \bibnamefont{Mei{\ss}ner}}, \bibinfo{journal}{Eur. Phys. J. A}
  \textbf{\bibinfo{volume}{26}}, \bibinfo{pages}{107} (\bibinfo{year}{2005}).

\bibitem[{\citenamefont{Tarasov et~al.}(2011)\citenamefont{Tarasov, Briscoe,
  Gao, Kudryavtsev, and Strakovsky}}]{PhysRevC.84.035203}
\bibinfo{author}{\bibfnamefont{V.~E.} \bibnamefont{Tarasov}},
  \bibinfo{author}{\bibfnamefont{W.~J.} \bibnamefont{Briscoe}},
  \bibinfo{author}{\bibfnamefont{H.}~\bibnamefont{Gao}},
  \bibinfo{author}{\bibfnamefont{A.~E.} \bibnamefont{Kudryavtsev}},
  \bibnamefont{and} \bibinfo{author}{\bibfnamefont{I.~I.}
  \bibnamefont{Strakovsky}}, \bibinfo{journal}{Phys. Rev. C}
  \textbf{\bibinfo{volume}{84}}, \bibinfo{pages}{035203}
  (\bibinfo{year}{2011}).

\bibitem[{\citenamefont{Landau and Lifshits}(1965)}]{LandauBook}
\bibinfo{author}{\bibfnamefont{L.~D.} \bibnamefont{Landau}} \bibnamefont{and}
  \bibinfo{author}{\bibfnamefont{E.~M.} \bibnamefont{Lifshits}},
  \emph{\bibinfo{title}{Quantum Mechanics : Non-Relativistic Theory}}
  (\bibinfo{publisher}{Pergamon Press}, \bibinfo{year}{1965}).

\bibitem[{\citenamefont{Levchuk et~al.}(2006)\citenamefont{Levchuk, Loginov,
  Sidorov, Stibunov, and Schumacher}}]{PhysRevC.74.014004}
\bibinfo{author}{\bibfnamefont{M.~I.} \bibnamefont{Levchuk}},
  \bibinfo{author}{\bibfnamefont{A.~Y.} \bibnamefont{Loginov}},
  \bibinfo{author}{\bibfnamefont{A.~A.} \bibnamefont{Sidorov}},
  \bibinfo{author}{\bibfnamefont{V.~N.} \bibnamefont{Stibunov}},
  \bibnamefont{and}
  \bibinfo{author}{\bibfnamefont{M.}~\bibnamefont{Schumacher}},
  \bibinfo{journal}{Phys. Rev. C} \textbf{\bibinfo{volume}{74}},
  \bibinfo{pages}{014004} (\bibinfo{year}{2006}).

\bibitem[{\citenamefont{D\"oring et~al.}(2004)\citenamefont{D\"oring, Oset, and
  Vicente~Vacas}}]{PhysRevC.70.045203}
\bibinfo{author}{\bibfnamefont{M.}~\bibnamefont{D\"oring}},
  \bibinfo{author}{\bibfnamefont{E.}~\bibnamefont{Oset}}, \bibnamefont{and}
  \bibinfo{author}{\bibfnamefont{M.~J.} \bibnamefont{Vicente~Vacas}},
  \bibinfo{journal}{Phys. Rev. C} \textbf{\bibinfo{volume}{70}},
  \bibinfo{pages}{045203} (\bibinfo{year}{2004}).

\bibitem[{\citenamefont{Machleidt}(2001)}]{PhysRevC.63.024001}
\bibinfo{author}{\bibfnamefont{R.}~\bibnamefont{Machleidt}},
  \bibinfo{journal}{Phys. Rev. C} \textbf{\bibinfo{volume}{63}},
  \bibinfo{pages}{024001} (\bibinfo{year}{2001}).

\bibitem[{\citenamefont{Bernard et~al.}(1996)\citenamefont{Bernard, Kaiser, and
  Mei{\ss}ner}}]{BERNARD1996116}
\bibinfo{author}{\bibfnamefont{V.}~\bibnamefont{Bernard}},
  \bibinfo{author}{\bibfnamefont{N.}~\bibnamefont{Kaiser}}, \bibnamefont{and}
  \bibinfo{author}{\bibfnamefont{U.-G.} \bibnamefont{Mei{\ss}ner}},
  \bibinfo{journal}{Phys. Lett. B} \textbf{\bibinfo{volume}{383}},
  \bibinfo{pages}{116 } (\bibinfo{year}{1996}).

\end{thebibliography}

\end{document}